\pdfoutput=1

\documentclass[11pt]{article}

\usepackage[final]{acl}

\usepackage{times}
\usepackage{latexsym}

\usepackage[T1]{fontenc}

\usepackage[utf8]{inputenc}

\usepackage{microtype}

\usepackage{inconsolata}

\usepackage{graphicx}

\usepackage{pifont}

\usepackage{amsmath,amsfonts,amssymb}
\usepackage{bm}

\usepackage{multirow}
\usepackage{makecell}
\usepackage{subfigure}
\usepackage{booktabs}

\newcolumntype{H}{>{\setbox0=\hbox\bgroup}c<{\egroup}@{}}  

\newcommand{\x}{\bm{\mathrm{x}}}
\newcommand{\z}{\bm{\mathrm{z}}}
\newcommand{\txt}{\bm{\mathrm{t}}}

\title{PALM: Few-Shot Prompt Learning for Audio Language Models}

\author{Asif Hanif,~Maha Tufail Agro,~Mohammad Areeb Qazi,~Hanan Aldarmaki\\
        Mohamed Bin Zayed University of Artificial Intelligence (MBZUAI)\\
        \{\texttt{asif.hanif, maha.tufail, mohammad.qazi, hanan.aldarmaki\}@mbzuai.ac.ae}}

\usepackage{printlen}

\begin{document}
\maketitle
\begin{abstract}
Audio-Language Models (ALMs) have recently achieved remarkable success in zero-shot audio recognition tasks, which match features of audio waveforms with class-specific text prompt features, inspired by advancements in Vision-Language Models (VLMs). Given the sensitivity of zero-shot performance to the choice of hand-crafted text prompts, many prompt learning techniques have been developed for VLMs. We explore the efficacy of these approaches in ALMs and propose a novel method, \textit{Prompt Learning in Audio Language Models (PALM)}, which optimizes the feature space of the text encoder branch. Unlike existing methods that work in the input space, our approach results in greater training efficiency. We demonstrate the effectiveness of our approach on 11 audio recognition datasets, encompassing a variety of speech-processing tasks, and compare the results with three baselines in a few-shot learning setup.  Our method is either on par with or outperforms other approaches while being computationally less demanding. Code is available at \url{https://asif-hanif.github.io/palm/}. 
\end{abstract}


\section{Introduction}
Inspired by the success of Vision-Language Models (VLMs) \cite{zhang2024vision}, Audio-Language Models (ALMs) have recently emerged, achieving state-of-the-art performance on various zero-shot audio recognition tasks \cite{elizalde2023clap, deshmukh2023pengi, kong2024audio, das2024speechverse}. In zero-shot audio recognition, features of the audio waveform are matched with features of text prompts representing each class, and the highest matching class is assigned to the audio waveform.
Zero-shot audio recognition offers significant advantages by eliminating the need for extensive labeled datasets and allowing for the recognition of new classes without additional training. This approach reduces training times and data annotation costs, leading to substantial savings in computational resources.\\

\begin{figure*}[!t]
    \centering
    \includegraphics{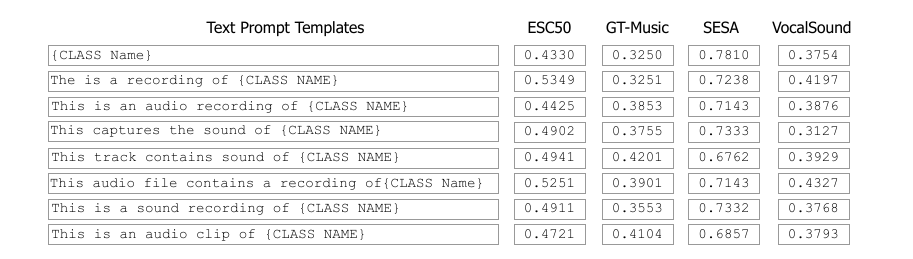}
    \caption{\textbf{Impact of Hand-crafted Prompts on ZERO-SHOT Performance} Zero-shot accuracy across four audio recognition datasets (ESC50 \cite{piczak2015dataset}, GT-Music-Genre \cite{sturm2012analysis}, SESA \cite{spadini2019sound}, and VocalSound \cite{gong_psla}) is evaluated with eight different text prompts using PENGI \cite{deshmukh2023pengi} model. The accuracy varies with changes in the handcrafted prompts.}
    \label{fig:text_prompt_templates}
\end{figure*}

The choice of text prompt is crucial for pre-trained vision-language and audio-language models, but it becomes a drawback for zero-shot recognition due to the requirement of hand-crafted prompts. This manual prompt-engineering can result in performance variations \cite{zhou2022coop, zhou2022conditional}. We confirm this observation, previously noted in VLMs, within the context of ALMs (refer to Figure \ref{fig:text_prompt_templates}). To automate the learning of text prompts, various approaches have been introduced for prompt learning in VLMs \cite{Gu2023ASS}. \\

\begin{figure}[t]
    \centering
    \includegraphics[width=\columnwidth]{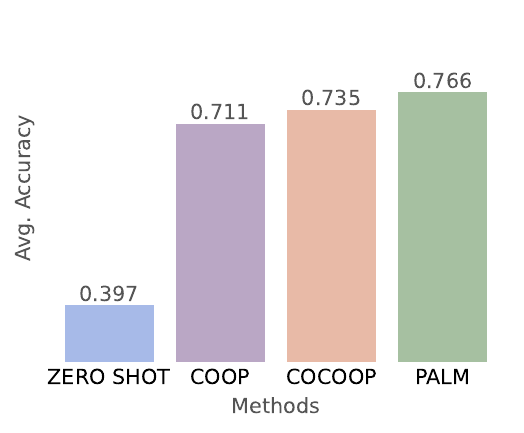}
    \caption{Comparison of our proposed approach, PALM, with three baselines: ZERO-SHOT \cite{deshmukh2023pengi}, COOP \cite{zhou2022coop} and COCOOP \cite{zhou2022conditional}. Bar plots show classification accuracy averaged across 11 audio datasets encompassing various speech-processing tasks.}
    \label{fig:avg_results_methods}
\end{figure}

The domain of prompt learning in ALMs remains under-explored, lacking comprehensive studies to evaluate the efficacy of prompt learning techniques within this context. To bridge this research gap, we adapt prompt learning techniques developed for VLMs and apply them to the domain of ALMs. Our results demonstrate that these adaptations improve the audio recognition performance (see Table \ref{tab:main_results}). Traditional techniques optimize the input space (token embeddings) of the text encoder branch by introducing a learnable context. However, this approach can increase training costs as loss gradients must flow through the text encoder branch. To address this, we introduce a novel method, \textbf{PALM}: \textbf{P}rompt Learning in \textbf{A}udio \textbf{L}anguage \textbf{M}odels, which optimizes the feature space of the text encoder rather than its input space. This makes the training computationally efficient since loss gradients do not need to flow through the text encoder. To assess the effectiveness of our approach,  we show results on 11 audio recognition datasets, encompassing various speech processing tasks. Our method either matches or surpasses other approaches while being less computationally demanding (see Table \ref{tab:main_results} and Table \ref{tab:methods_num_params}), setting a benchmark for prompt learning in ALMs.  \\

\noindent \textbf{Contributions}: Our contributions are as follows:
\begin{itemize}
    \renewcommand\labelitemi{--}
    \item Inspired by the success of few-shot based prompt learning in VLMs, we are the first (to the best of our knowledge) to demonstrate its efficacy in ALMs.
    \item We show that prompt learning techniques, initially developed for VLMs, can significantly enhance the performance when adapted for ALMs.
    \item We introduce a novel few-shot based prompt learning method, PALM, for ALMs that optimizes the feature space of the text encoder, outperforming existing baselines.
    \item We demonstrate our approach's effectiveness on 11 audio recognition datasets, comparing it to three baselines in a few-shot learning setup. Our method matches or outperforms others while being less computationally demanding, establishing a benchmark for prompt learning in audio-language models and paving the way for future research.
\end{itemize}

\section{Related Work}
Prompt engineering involves adding task-specific hints, called prompts, to a large pre-trained model to adapt it to new tasks. Recently, significant advancements have been made in prompt learning, particularly in the fields of language and vision. Below, we outline the recent developments in language, vision, and audio domains.
\subsection{Audio Language Models (ALMs)} 
Taking inspiration from multimodal models like CLIP \cite{radford2021learning} in the vision domain, Contrastive Language-Audio Pretraining (CLAP) \cite{elizalde2023clap} stands out as the first-of-its-kind audio language model. It connects natural language and audio through dual encoders and contrastive learning, aligning audio and text descriptions in a shared multimodal space. Furthermore, CLAP introduces zero-shot prediction capabilities, removing the necessity for training with predefined class labels and allowing flexible class prediction during inference.

PENGI \cite{deshmukh2023pengi}, another audio language Model, utilizes transfer learning by treating all audio tasks as text-generation tasks. It takes audio recordings and text inputs, generating free-form text as output. The input audio is represented by continuous embeddings from an audio encoder, while the corresponding text input undergoes the same process with a text encoder. These sequences are combined as a prefix to prompt a pre-trained frozen language model. PENGI's unified architecture supports both open-ended and close-ended tasks without requiring additional fine-tuning or task-specific extensions.

Audio Flamingo, introduced by \citet{kong2024audio}, is a multimodal-to-text generative model inspired by Flamingo \cite{NEURIPS2022_960a172b}, demonstrating advanced audio understanding capabilities, adaptability to unseen tasks through in-context learning and retrieval, and multi-turn dialogue abilities. The model features an audio feature extractor with a sliding window and uses cross-attention to fuse audio inputs into the language model, ensuring computational efficiency.

\subsection{Prompt Learning in Language Models}
Extensive research has been conducted on prompt learning techniques in natural language processing. Pioneering work by \cite{NEURIPS2020_1457c0d6} focused on optimization strategies for zero-shot and few-shot learning scenarios, demonstrating that prompts can enable generative models to perform well across various tasks without extensive task-specific training. Their method leverages the model's pre-trained knowledge and prompt-guided interactions to achieve strong performance on new tasks. They also introduced GPT-3, which transformed the field of prompt learning in natural language processing. \citet{petroni2019language} integrated contextual cues and constraints within prompts to guide model behavior, embedding task-specific information to enhance output precision and relevance. Their technique improves interpretability and task-oriented performance by providing contextual guidance during inference.

\subsection{Prompt Learning in Vision-Language Models}
Inspired by advancements in prompt-based work in language models, several studies have been conducted to adapt these methods to VLMs \cite{Gu2023ASS}. Some focus exclusively on the language component, such as COOP (Context Optimization) \cite{zhou2022conditional}. In contrast, others integrate insights from language and visual components, as seen in COCOOP (Conditional Context Optimization) \cite{zhou2022conditional}. COOP enhances CLIP model's few-shot transfer learning capability by optimizing a continuous set of prompt vectors within the language branch. However, COCOOP addresses the limitations of COOP, particularly its suboptimal performance on novel classes, by explicitly conditioning prompts on individual image instances, thereby enhancing generalization.

\subsection{Prompt Learning in Audio-Language Models} 
Prompt learning with audio-language models is relatively understudied. Previous work has explored enhancing language models with speech recognition by conditioning them on variable-length audio embeddings using a conformer-based audio encoder \cite{fathullah2024prompting}. \citet{deshmukh2024domain} propose a test-time domain adaptation method for Contrastive ALMs, using unlabeled audio to adjust the model to new domains via a domain vector, consistent predictions, and self-entropy fine-tuning, improving on traditional Test-Time Training. \citet{li2024audio} introduce \textit{PT-Text}, an audio-free prompt tuning scheme that optimizes prompt tokens from text, regularizing the model to avoid overfitting by training with captions and using a multi-grained strategy to enhance performance. Despite these advancements, more research is needed to fully understand and exploit prompt learning in audio-language models.

\section{Method}
\subsection{Audio-Language Model (ALM)}
\label{ss:intro_alm}
We demonstrate the efficiency of prompt learning in enhancing zero-shot performance using a state-of-the-art audio-language model PENGI \cite{deshmukh2023pengi}. Our approach is applicable to all audio-language models that have \textit{aligned} audio and text encoders.\\\\
\textbf{PENGI} takes an audio waveform and a text prompt as input and generates free-form text. It consists of three branches. The first branch is an audio encoder that maps the audio waveform to an embedding space. The second branch is a text encoder that transforms the input text into the same embedding space. These embeddings are then concatenated to form an input prefix for the third branch, a causal language model that generates tokens autoregressively, conditioned on both the audio and text inputs. PENGI can be used for various audio-conditioned tasks, such as text completion, classification, audio caption generation, and question-answering \cite{deshmukh2023pengi}. 
\\\\
\textbf{Zero-Shot Inference} Although PENGI is multimodal-to-text generation model, however, we use its audio and text encoder branches for zero-shot audio recognition.  This is accomplished by comparing the embedding of the audio waveform (extracted from the audio encoder) with the embeddings of text prompts for different classes (extracted from the text encoder). An overview of zero-shot inference is given in Figure \ref{fig:main_diagram}(a). The zero-shot setup used by \cite{deshmukh2023pengi} differs from ours, as they employ the model's free-form text output for zero-shot inference. \\\\
Formally, we denote the pre-trained ALM as $f_{_{\theta}} = \{f_{_{A}},f_{_{T}}\}$, whereas $f_{_{A}}$ and $f_{_{T}}$ are audio and text encoders, respectively and $\theta$ represents the combined weights of both encoders. 
For classification in zero-shot scenario, an audio waveform $\x$ is first passed to the audio encoder $f_{_{A}}$ to produce a $d-$dimensional feature vector $f_{_{A}}(\x) \in \mathbb{R}^{d}$. In parallel, text prompts representing each class label $y_i \in \{y_1, y_2 \dots, y_c \}$ are encapsulated within class-specific handcrafted text templates, such as
$$t_i = \text{``}\texttt{An audio recording of}~\{\mathrm{CLASS}~y_i\}\text{''},$$ 
where $c$ is the total number of classes. Each text prompt, represented as $t_i$, is processed through the text encoder $f_{_{T}}$, resulting in a feature vector $f_{_{T}}(t_i) \in \mathbb{R}^{d}$. The relationship between the audio waveform $\x$ and a class-specific text prompt $t_i$ is quantified by computing the cosine similarity between their corresponding feature vectors, denoted as $\mathtt{sim}(f_{_{A}}(\x)~,~f_{_{T}}(t_i))$. The class with the highest similarity score is then assigned as the label $\hat{y}$ for the audio waveform, i.e.
\begin{equation}
\label{eq:zeroshot_infer}
\hat{y} = \underset{ i\in \{1,2,\dots,c\} }{\mathbf{argmax}} ~~~ \mathtt{sim}\big(f_{_{A}}(\x)~,~f_{_{T}}(t_i)\big). 
\end{equation}

\subsection{PALM: Prompt Learning in ALM}
\label{ss:palm}
\begin{figure*}[!th]
    \centering
    \frame{\includegraphics[trim={0.3in 0in 0.4in 0in},clip,scale=1.05]{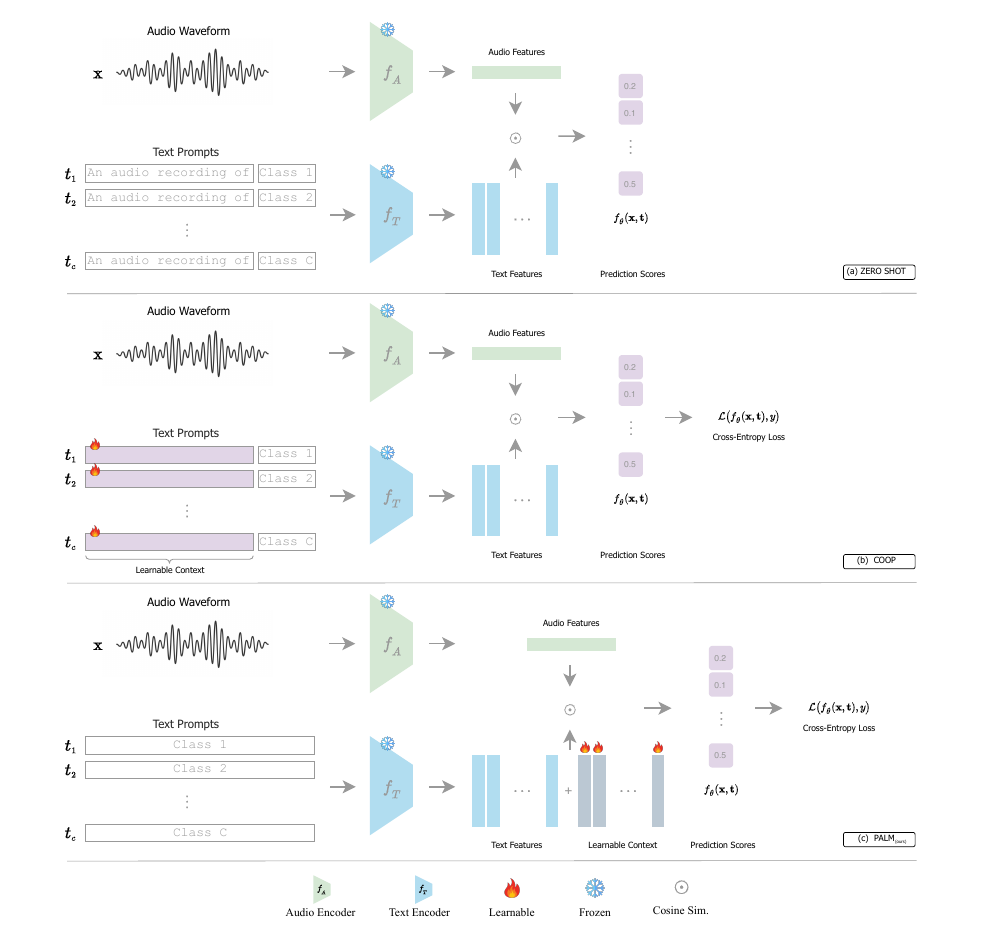}}
    \caption{\textbf{Overview of Zero-Shot, COOP, PALM} \textit{(a)} \textbf{Zero-Shot} inference involves matching the embedding of the audio waveform with the embeddings of text prompts for each class. The class with the highest matching score is then assigned to the audio. \textit{(b)}  \textbf{COOP} \cite{zhou2022coop} avoids using handcrafted text prompts by learning the context of text prompts in the token embedding space. It optimizes the input space of the text encoder to enhance classification performance. \textit{(c)} \textbf{PALM} requires only class names at the input of text encoder and it optimizes the feature space by adding learnable context embeddings to text feature vectors. PALM not only outperforms COOP (see Table \ref{tab:main_results}), but it is also more computationally efficient since it does not require gradients to flow through the text encoder, unlike COOP.}
    \label{fig:main_diagram}  
\end{figure*}

In our proposed method, we do not use hand-crafted prompts; instead, we simply use class names as the input to the text encoder i.e. $t_i = \text{``}\{\mathrm{CLASS}~y_i\}\text{''}$. Moreover, unlike COOP \cite{zhou2022coop}, which learns the context of input text prompts in the token embedding space (see Figure \ref{fig:main_diagram}(b)), we learn the context in the feature space of prompts. Specifically, after obtaining the feature vector of the $i_{\text{th}}$ class text prompt via the text encoder, i.e., $f_{_{T}}(t_i) \in \mathbb{R}^{d}$, we add a learnable vector $z_i \in \mathbb{R}^{d}$ to it to get the updated text feature vector as follows:
\begin{equation}
\label{eq:update_ft}
f_{_{T}}^{\prime}(t_i) = (1-\lambda_i)\cdot f_{_{T}}(t_i)~+~\lambda_i \cdot z_i   
\end{equation}
where $\lambda_i \in [0,1]$ is a learnable parameter that determines the contributions of both vectors. Assuming 
$\txt=\{t_1,t_2,\dots,t_c\}$ denotes text prompts of all classes, the raw/un-normalized prediction scores (logits), denoted as $f_{_{\theta}}(\x,\mathbf{t}) \in \mathbb{R}^{c}$, for an audio waveform $(\x)$ are obtained as follows:
$$f_{_{\theta}}(\x,\txt) = \bigg\{~\mathtt{sim}\bigg(f_{_{A}}(\x)~,~f_{_{T}}^{\prime}(t_i)\bigg)~\bigg\}_{i=1}^{c},$$
where $\texttt{sim}(\cdot)$ is cosine-similarity function and $c$ is the number of classes. $f_{_{A}}(\x)$ is the feature vector from the audio encoder, and  $f_{_{T}}^{\prime}(t_i)$ is the updated text feature vector (Equation \ref{eq:update_ft}) of $i_{\text{th}}$ class.\\\\
We optimize the following objective function to learn feature-space context embeddings $\z=\{z_1,z_2,\dots,z_c\}$ and their corresponding contributions $\lambda=\{\lambda_1,\lambda_2,\dots,\lambda_c\}$,
\begin{equation}
\label{eq:palm_objective}
\underset{ \z~,~\lambda }{\mathbf{minimize}}~~ \sum_{(\x,y)\in\mathcal{D}} \mathcal{L}\big(f_{_{\theta}}(\x,\txt),y\big),
\end{equation} 
where $\mathcal{D}=\{\x_i,y_i\}_{i=1}^{N}$ is training dataset consisting of $N$ audio-class pairs and $\mathcal{L}(\cdot)$ denotes cross-entropy loss. We use  \textit{few-shot} setting during training, meaning that a fixed number of samples (e.g., 16) are randomly selected from each class in the training dataset.  While optimizing objective in Equation \ref{eq:palm_objective}, weights of both encoders $\{f_{_{A}},f_{_{T}}\}$ are kept in frozen state. The number of learnable parameters in our proposed method is $c+(c\times d)$. After learning the parameters, we use Equation \ref{eq:palm_infer} for audio classification during inference stage.
\begin{equation}
\label{eq:palm_infer}
\hat{y} = \underset{ i\in \{1,2,\dots,c\} }{\mathbf{argmax}} ~~~ \mathtt{sim}\big(f_{_{A}}(\x)~,~f_{_{T}}^{\prime}(t_i)\big) 
\end{equation}
An overview of our proposed approach can be found in Figure \ref{fig:main_diagram}(c).

\subsection{Difference with COOP and COCOOP}
\label{ss:diff_palm_coop}
COOP \cite{zhou2022coop} and COCOOP \cite{zhou2022conditional} were originally introduced for vision-language model; however, we adapted them for audio-language model (replacing the vision encoder branch with audio encoder branch) and presented it as baseline methods. Both of these baselines and our method aim to enhance zero-shot performance for audio classification in this work. While our method and the baselines share this common goal, they differ in their approach to achieving it. COOP and COCOOP optimize the input space (token embeddings of prompt context) of text encoder, whereas our method optimizes the text feature space. In our method, loss gradients do not need to flow through the text encoder, whereas in COOP and COCOOP, gradients flow through the encoder to reach the input to update prompt context. Moreover, there is a feedback loop from audio features (output of audio encoder) to the input of text encoder in COCOOP, making it even more computationally expensive. Comparatively, our method is more computationally efficient as it does not include a feedback loop (see Table \ref{tab:methods_num_params}). Both COOP and COCOOP require a user-specified hyper-parameter, namely the number of context tokens, whereas our method does not rely on such a parameter. Results in Table \ref{tab:main_results} demonstrate that our method outperforms COOP and COCOOP, achieving an average improvement of 5.5\% and 3.1\% respectively.

\section{Experiments and Results}
\subsection{Datasets}
\begin{table}[]
    \centering
    \scalebox{0.59}{%
    \begin{tabular}{l|c|c|c}
    \toprule
    $\bm{\mathrm{DATASETS}}$ & $\bm{\mathrm{TYPE}}$ & $\bm{\mathrm{CLASSES}}$ & $\bm{\mathrm{SPLIT}}$  \\
    \midrule \midrule
    Beijing-Opera & \multirow{2}{*}{Instrument Classification} & 4 & Five Fold \\
    NS-Instruments &  & 10 &  Train-Test \\
    \midrule
    ESC50 & \multirow{3}{*}{Sound Event Classification} & 50 & Five Fold \\
    ESC50-Actions &  & 10 & Five Fold \\
    UrbanSound8K &  & 10 & Ten Fold \\
    \midrule
    CREMA-D & \multirow{2}{*}{Emotion Recognition} & 6 & Train-Test \\
    RAVDESS &  & 8 & Train-Test \\
    \midrule
    VocalSound & Vocal Sound Classification & 6 & Train-Test \\
    \midrule
    SESA & Surveillance Sound Classification & 4 & Train-Test \\
    \midrule
    TUT2017 & Acoustic Scene Classification & 15 & Four Fold \\
    \midrule
    GT-Music-Genre & Music Analysis & 10 & Train-Test  \\
    \bottomrule
    \end{tabular}
    }
    \caption{\textbf{Datasets Information} In this work, we use 11 multi-class classification datasets encompassing a variety of speech-processing tasks.}
    \label{tab:dataset_stats}
\end{table}

We evaluate our methodology using datasets from various speech-processing tasks: instrument classification, sound event classification, emotion recognition, vocal sound classification, surveillance sound event classification, acoustic scene classification, and music analysis. Brief information of each dataset can be found in Table \ref{tab:dataset_stats}. For instrument classification, we use Beijing-Opera \cite{tian2014study} dataset, which includes audio examples of strokes from four percussion instrument classes used in Beijing Opera, and NS-Instruments \cite{nsynth2017} dataset, which consists of one-shot instrumental notes with unique pitches, timbres, and envelopes, spanning ten classes. For sound event classification, we utilize three datasets: ESC50 \cite{piczak2015dataset}, containing environmental recordings across 50 classes; ESC50-Actions \cite{piczak2015dataset}, a subset with 10 classes of non-speech human sounds; and UrbanSound8K \cite{salamon2014dataset}, with urban noise excerpts from 10 classes. Emotion recognition is assessed with the CREMA-D \cite{cao2014crema} and RAVDESS \cite{livingstone2018ryerson} datasets, covering 6 and 8 emotion classes respectively, performed by actors. We employ the VocalSound \cite{gong_psla} dataset for vocal sound classification, which includes 6 classes of human non-speech vocalizations. For surveillance sound event classification, we use SESA \cite{spadini2019sound} dataset, which has 4 classes. Acoustic scene classification uses the TUT2017 \cite{heittola2017tut} dataset, containing samples from 15 acoustic scenes. For music analysis, the GT-Music-Genre \cite{sturm2012analysis} dataset is used, which includes 10 classes of music genres.\\ 

\noindent We adhere to the official train-test or multi-fold splits for all datasets. We conduct cross-validation experiments on datasets having multi-fold splits such as Beijing-Opera, ESC50, ESC50-Actions, UrbanSound8K, and TUT2017, and report the average scores. We have publicly released all information regarding dataset preprocessing to ensure reproducibility of results.

\subsection{Baseline Methods}
For baselines, we consider PENGI model \cite{deshmukh2023pengi} (in ZERO-SHOT setup), COOP \cite{zhou2022coop} and COCOOP \cite{zhou2022conditional}. COOP and COCOOP are prompt learning approaches, originally introduced for VLMs. Both of these approaches remove the requirement of providing handcrafted text prompts and they optimize the input token embedding space of text encoder to enhance accuracy. The only difference between COOP and COCOOP is that the latter incorporates a feedback loop from the output of the audio encoder to the input of the text encoder. We adapt these two approaches for audio-language models by replacing the vision encoder with an audio encoder and present them as baselines for our proposed method.
\textit{Why PENGI, COOP and COCOOP as baselines?} PENGI is an state-of-the-art ALM that has demonstrated comprehensive evaluation across 21 downstream audio tasks, making it a robust benchmark for comparison. COOP and COCOOP, on the other hand, are pioneering works on prompt learning in the domain of vision-language models, offering foundational techniques and insights that are highly relevant for our study. 

\begin{table*}[!t]
    \centering
    \setlength{\tabcolsep}{4pt}
    \scalebox{0.67}{%
    \begin{tabular}{l|c|cccc|cccc|cccc}
    \toprule
    $\bm{\mathrm{METHODS}}$ $\rightarrow$ & \multicolumn{1}{|c}{$\bm{\mathrm{ZERO~SHOT}}$} & \multicolumn{4}{|c}{$\bm{\mathrm{COOP}}$} & \multicolumn{4}{|c}{$\bm{\mathrm{COCOOP}}$} & \multicolumn{4}{|c}{$\bm{\mathrm{PALM}}_{\mathrm{(ours)}}$} \\
    \cmidrule(lr{3pt}){2-2} \cmidrule(lr{3pt}){3-6} \cmidrule(lr{3pt}){7-10} \cmidrule(lr{3pt}){11-14} 
    $\bm{\mathrm{DATASETS}}$ $\downarrow$ & - & SEED-0 & SEED-1 & SEED-2 & AVG & SEED-0 & SEED-1 & SEED-2 & AVG & SEED-0 & SEED-1 & SEED-2 & AVG  \\
    \midrule \midrule
    Beijing-Opera & 0.2881 & 0.9323 & 0.9660 & 0.9619 & 0.9534 & 0.9577 & 0.9830 & 0.9916 & \textbf{0.9774} & 0.9747 & 0.9066 & 0.9787 & 0.9533 \\
    CREMA-D & 0.2310 & 0.3130 & 0.4197 & 0.2760 & 0.3362 & 0.2539 & 0.3358 & 0.3156 & 0.3018 & 0.4453 & 0.3580 & 0.2344 & \textbf{0.3459} \\
    ESC50-Actions & 0.6525 & 0.9625 & 0.9400 & 0.9550 & 0.9525 & 0.9631 & 0.9620 & 0.9648 & 0.9634 & 0.9700 & 0.9625 & 0.9650 & \textbf{0.9658} \\
    ESC50 & 0.4965 & 0.9410 & 0.9390 & 0.9345 & 0.9382 & 0.9460 & 0.9370 & 0.9450 & 0.9427 & 0.9560 & 0.9600 & 0.9620 & \textbf{0.9593} \\
    GT-Music-Genre & 0.3250 & 0.7250 & 0.6950 & 0.7350 & 0.7183 & 0.7500 & 0.7450 & 0.7600 & 0.7517 & 0.7900 & 0.7850 & 0.8250 & \textbf{0.8000} \\
    NS-Instruments & 0.3291 & 0.5728 & 0.5562 & 0.6177 & 0.5822 & 0.5996 & 0.5740 & 0.6438 & 0.6058 & 0.6394 & 0.6108 & 0.6648 & \textbf{0.6383} \\
    RAVDESS & 0.1222 & 0.3849 & 0.2688 & 0.3422 & 0.3320 & 0.3727 & 0.4399 & 0.3523 & 0.3883 & 0.4562 & 0.4603 & 0.4623 & \textbf{0.4596} \\
    SESA & 0.7238 & 0.9143 & 0.8952 & 0.8762 & 0.8952 & 0.8381 & 0.8762 & 0.8952 & 0.8698 & 0.8857 & 0.9143 & 0.8857 & \textbf{0.8952} \\
    TUT2017 & 0.2435 & 0.6391 & 0.6667 & 0.6525 & 0.6528 & 0.7499 & 0.7215 & 0.7312 & 0.7342 & 0.7959 & 0.8047 & 0.7729 & \textbf{0.7912} \\
    UrbanSound8K & 0.5349 & 0.7600 & 0.7378 & 0.7666 & 0.7548 & 0.7576 & 0.7784 & 0.7597 & 0.7652 & 0.8120 & 0.8037 & 0.8074 & \textbf{0.8077} \\
    VocalSound & 0.4197 & 0.7162 & 0.7485 & 0.6642 & 0.7096 & 0.8081 & 0.7825 & 0.7463 & 0.7790 & 0.8101 & 0.8168 & 0.7964 & \textbf{0.8078} \\
    \midrule
    AVERAGE & 0.3969 & 0.7146 & 0.7121 & 0.7074 & 0.7114 & 0.7276 & 0.7396 & 0.7369 & 0.7347 & 0.7759 & 0.7621 & 0.7595 & \textbf{0.7658} \\
    \bottomrule
    \end{tabular}
    }
    \caption{\textbf{Comparison of $\mathrm{PALM}$ with Baselines} The accuracy scores of the methods (ZERO SHOT \cite{deshmukh2023pengi}, COOP \cite{zhou2022coop}, COCOOP \cite{zhou2022conditional}, and our proposed method PALM) across 11 datasets are presented. For each method (except ZERO SHOT), experiments were performed using three different seeds. The accuracy scores for all seeds are reported, along with the average score. Bold values indicate the best average score in each row. Compared to the baselines, our proposed method achieves favorable results, with an average improvement of 5.5\% over COOP and 3.1\% over COCOOP. It should be noted that both COOP and COCOOP are computationally expensive, as these approaches require loss gradients to flow through the text encoder. Additionally, COCOOP has a feedback loop from audio features to the input space of the text encoder, making it even more computationally expensive. On the other hand, PALM is relatively less computationally expensive.}
    \label{tab:main_results}
\end{table*}

\subsection{Experimental Setup}
We use pre-trained PENGI \cite{deshmukh2023pengi} as the audio-language model for all methods. For all methods, except ZERO-SHOT, we conduct experiments for $50$ epochs. Following the few-shot evaluation setup, we use 16 randomly selected samples per class from the training dataset. For inference, we utilize the entire test dataset. In the case of multi-fold datasets, we employ cross-validation and report the average scores. Training is performed using the Stochastic Gradient Descent (SGD) optimizer with a learning rate of $0.05$. We use `Accuracy' as the evaluation metric. For all methods, except ZERO-SHOT, we run experiments with three different seeds and report the scores for each seed along with the average score. For ZERO-SHOT, we use default text prompt template ``\texttt{This is a recording of \{CLASS NAME\}}". For COOP \cite{zhou2022coop} and COCOOP \cite{zhou2022conditional} baselines, we set the number context tokens to $16$ and context is placed at the front of class names. PENGI model weights are kept ``frozen'' in all experiments. We use \texttt{NVIDIA A100-SXM4-40GB} GPU for all experiments and Pytorch version \texttt{1.11+cuda11.3}. 

\subsection{Results}
Table \ref{tab:main_results} presents the performance comparison across 11 datasets using four different methods. Results indicate that PALM generally outperforms COOP and COCOOP, showing an average improvement of 5.5\% over COOP and 3.1\% over COCOOP. This suggests that PALM is a more effective approach in most cases. Moreover, it is important to note that PALM uses significantly fewer parameters—87\% fewer compared to COCOOP. This reduction in parameters can contribute to more efficient model training and deployment. 

\begin{table}[]
    \centering
    \scalebox{0.65}{%
    \begin{tabular}{l|c|c|c|c}
    \toprule
    $\bm{\mathrm{METHOD}}$ & ${\mathrm{ZERO~SHOT}}$ & ${\mathrm{COOP}}$ & ${\mathrm{COCOOP}}$ & ${\mathrm{PALM}}$ \\
    \midrule
    $\bm{\mathrm{\#~of~Parameters}}$ & 0 & $8,192$ & $98,880$ & $12,393$ \\
    \bottomrule
    \end{tabular}
    }
    \caption{\textbf{Number of Learnable Parameters} in baselines and PALM.} 
    \label{tab:methods_num_params}
\end{table}
\begin{table}[]
    \centering
    \scalebox{0.65}{%
    \begin{tabular}{l|c}
    \toprule
    $\bm{\mathrm{METHOD}}$ & $\bm{\mathrm{AVERAGE~~ ACCURACY}}$ \\
    \midrule
    ${\mathrm{PALM+COOP}}$ & 0.7236  \\
    ${\mathrm{PALM+COCOOP}}$ & 0.7094\\
    ${\mathrm{PALM+COCOOP^{\dagger}}}$ & 0.7352 \\
    ${\mathrm{LINEAR~~PROBING}}$ & 0.7299  \\
    ${\mathrm{PALM}}^{\dagger}$ & 0.7160 \\
    ${\mathrm{PALM}}$ & \textbf{0.7658} \\
    \bottomrule
    \end{tabular}
    }
    \caption{\textbf{PALM+Baselines}  Jointly optimizing input and output space of text encoder does not help attain better accuracy. $\mathrm{COCOOP^{\dagger}}$ refers to the method where feedback from audio features is incorporated into the text features, rather than being fed directly into the text encoder's input. ${\mathrm{PALM}}^{\dagger}$ denotes experiment in which text features are not used.} 
    \label{tab:ablation}
\end{table}

In the datasets, namely Beijing-Opera, ESC50 and ESC50-Actions, the improvement of PALM over COCOOP is marginal. However, for the subsequent datasets, such as CREMA-D, GT-Music-Genre, NS-Instruments, RAVDESS, SESA, TUT2017, UrbanSound8K and VocalSound 
, the performance improvements are more substantial. This indicates that while PALM provides consistent benefits, its advantages become more pronounced with certain datasets.

\begin{figure*}[th]
    \centering
    \includegraphics[width=\textwidth, trim={0in 0.12in 0in 0.12in},clip,]{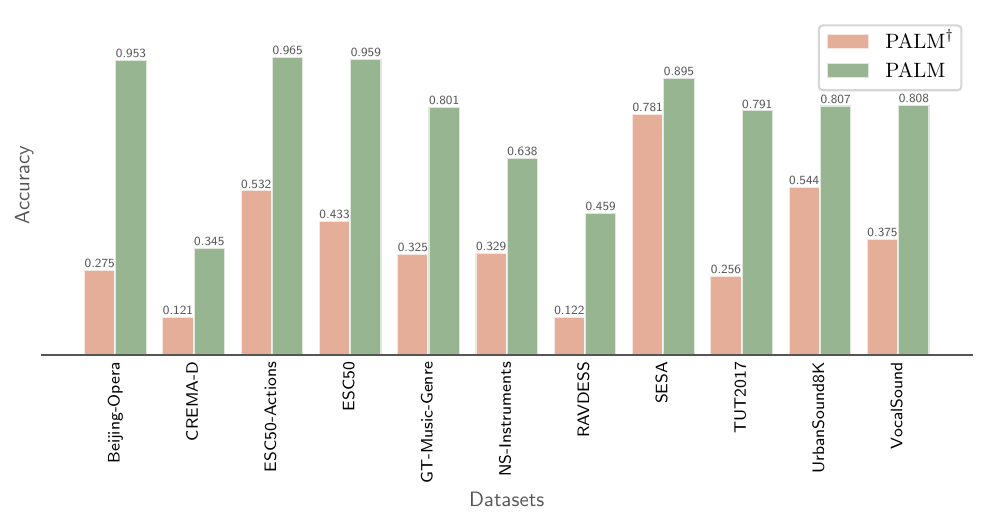}
    \caption{\textbf{Comparison of $\mathrm{PALM}^{\dagger}$ and $\mathrm{PALM}$}. Here, $\mathrm{PALM}^{\dagger}$ refers to setting in which the \textit{Learnable Context} embeddings (see Figure \ref{fig:main_diagram} for reference) have been \textbf{removed} from the feature space of the text encoder. The removal of context embeddings drastically degrades performance, highlighting their importance.}
    \label{fig:palm_vs_palm_no_context}
\end{figure*}
\begin{figure*}[h]
    \centering
    \includegraphics[width=\textwidth]{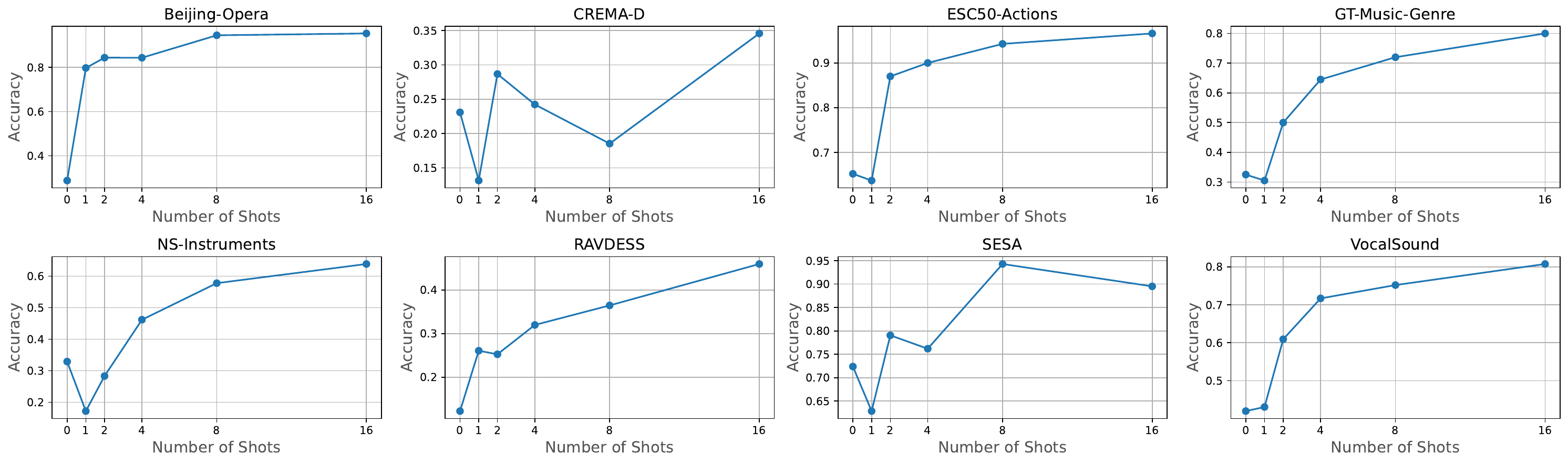}
    \caption{A higher number of shots generally leads to increased audio classification accuracy using PALM.}
    \label{fig:palm_num_shots}
\end{figure*}

\subsection{Ablative Analysis}
For ablative analysis, we first show the importance of incorporating learnable context embeddings in text features. In Figure \ref{fig:palm_vs_palm_no_context}, we compare the performance of our method with and without the learnable context embeddings. The results clearly demonstrate that removing the learnable context embeddings leads to a significant drop in performance, underscoring their crucial role in enhancing the model's accuracy. This highlights the effectiveness of our approach in optimizing the feature space of the text encoder. 

We also show the impact of jointly optimizing the input space and output space of the text encoder by applying PALM on top of COOP and COCOOP in Table \ref{tab:ablation}. The results indicate that joint optimization results in slight performance degradation and is not beneficial. Moreover, we also show linear probing results in Table \ref{tab:ablation}. Since our approach is based on few-shot setup, therefore, we show impact of number of shots (number of training samples per class) on the PALM's performance across eight datasets in Figure \ref{fig:palm_num_shots}. As the number of shots in the training dataset increases, the performance of the model tends to improve.

\section{Conclusion}
In this study, we investigate the application of prompt learning techniques, originally developed for vision-language models (VLMs), in the context of audio-language models (ALMs). We introduce PALM, a novel method that optimizes the feature space of the text encoder branch, enhancing training efficiency compared to existing methods that operate in the input space. Evaluated on 11 diverse audio recognition datasets, PALM consistently matches or surpasses established baselines in a few-shot learning setup while reducing computational demands. PALM offers a promising direction for enhancing the performance of ALMs in zero-shot and few-shot learning scenarios, contributing to the broader field of audio recognition and paving the way for future research in multimodal tasks.

\section*{Limitations}
Although we are the first, to the best of our knowledge, to integrate prompt learning techniques originally designed for Vision-Language Models (VLMs) into Audio-Language Models (ALMs) and propose a new method, several aspects still need to be addressed. One critical aspect is to analyze prompt learning performance for domain generalization. This involves evaluating how well the prompts adapt to new, unseen domains and tasks, ensuring robustness and effectiveness across various applications. The second aspect is to analyze prompt learning performance under different types of perturbations in audio data to check its resilience against various types of noise. This analysis is essential for understanding the robustness of the models in real-world scenarios where audio data can be contaminated with background noise, distortions, and other audio artifacts. Thirdly, while our study shows results on audio classification, it is yet to be seen how prompt learning helps in other audio tasks such as speech recognition, audio segmentation, and information retrieval. Investigating the effectiveness of prompt learning across a broader range of audio tasks will provide a more comprehensive understanding of its potential and limitations.

\bibliography{refs}

\end{document}